\documentclass[aps,pra,epsfigure,twocolumn,showpacs]{revtex4}
\usepackage{dcolumn}    
\usepackage{bm} 
\usepackage{graphicx}
\usepackage{amsmath}    
\usepackage{latexsym}
\usepackage{amsfonts}   
\usepackage{amssymb}
\usepackage{array}      
\usepackage{epsfig}

\newcommand{\ket}[1]{\left\vert#1\right\rangle}
\newcommand{\miniket}[1]{\vert#1\rangle}

\newcommand{\bra}[1]{\left\langle#1\right\vert}

\newcommand{\minisand}[3]{\langle#1\vert#2\vert#3\rangle}

\newcommand{\nbar}{\overline{n}}

\begin{document}

\title{Control-limited perfect state transfer, quantum stochastic resonance and many-body entangling gate in imperfect qubit registers}

\author{C. Di Franco$^1$, M. Paternostro$^1$, D. I. Tsomokos$^2$, and S. F. Huelga$^2$}

\affiliation{$^1$School of Mathematics and Physics, Queen's University, Belfast BT7 1NN, United Kingdom \\
$^2$ Quantum Physics Group, STRI, School of Physics, Astronomy \&
Mathematics, University of Hertfordshire, Hatfield, AL10
9AB, United Kingdom}

\begin{abstract}
We propose a protocol for perfect quantum state transfer that is resilient to a broad class of realistic experimental imperfections, including noise sources that could be modelled either as independent Markovian baths or as certain forms of  spatially correlated environments. We highlight interesting connections between the fidelity of state transfer and quantum stochastic resonance effects. The scheme is flexible enough to act as an effective entangling gate for the generation of genuine multipartite entanglement in a control-limited setting. Possible experimental implementations using superconducting qubits are also briefly discussed.
\end{abstract}

\date{\today}
 \maketitle

\section{Introduction}
\label{Intro}

In the last few years, it has been realized that specific forms of built-in and
permanent intra-register couplings {could} be used for the
purposes of quantum computation and
communication~\cite{generale,alwayson}. These methods allow
{us} to bypass the need for both fast and accurate
inter-qubit switching and gating, which are generally very
demanding tasks. We refer to such schemes -- where the required
external control over physical systems is greatly reduced -- as
{\it control-limited}. However, the price to pay for the
performance of efficient operations is the need to pre-engineer
appropriate patterns of couplings. Determining the exact
distribution of coupling strengths for a given interaction model
in a control-limited setting is often a matter of craftsmanship or
the result of the exploitation of certain geometrical properties
of the system at hand~\cite{cambridge,unmodulated}. A simple physical model
suitable for this scenario is provided by a linear spin chain.
Here we use the term ``spin chain" in its wider sense which
includes other physical systems as well that can be modelled by a
generic spin chain Hamiltonian.

In the quest for the realization of a realistic quantum processor,
the achievement of faithful transmission of quantum information
has been the object of remarkable interest. The possibility of
linking different local nodes of a quantum network through photons
has been exhaustively analyzed~\cite{flying}. For short-distance
quantum communication, the idea of using spin chains as {\it
quantum wires} has been put forward by Bose~\cite{unmodulated}.
With an isotropic Heisenberg interaction a transmission fidelity
that exceeds the maximum value achievable classically can be
obtained for a chain {of} up to $\sim80$ qubits. The original idea has then been 
extended along various directions~\cite{espanso}. In particular, Christandl
{\it et al.}~\cite{cambridge} showed that, by engineering the
strength of the couplings in the chain, {perfect state
transfer could be achieved}.

At the same time, quantum entanglement (bi-partite as well as
multi-partite) has been studied in great detail over the last
years~\cite{generale,entanglement} and appears to be a key
resource for many applications in quantum information processing
(QIP). It is well-known that genuine multipartite
entanglement of Greenberger-Horne-Zeilinger (GHZ) form~\cite{ghz}
is useful for multi-agent protocols of distributed QIP such as
quantum secret sharing, remote implementation of unknown
operations, quantum
average estimation and quantum anonymous transmission~\cite{qss,average}. Our study 
is thus distinguished, for instance, with respect to the investigation 
in~\cite{analogouscircuit}, which was centered on cluster state generation 
in quantum spin chains.

Here we show how to make use of a physical system whose
Hamiltonian can be mapped into that of a spin chain in order to
achieve both these goals: With the same system we are able to
transfer the state from one qubit to another or to generate
GHZ entanglement. In addition to
pragmatic goals relevant to quantum information processing, our
study reveals an interesting non-monotonic behavior of
state-transfer fidelity against the {strength} of the
external noise. This reminds us of quantum stochastic resonances in
many-body systems~\cite{susanamartin} and represents an original
way of revealing these fundamentally interesting
{processes} in a linear register of qubits. A possible
physical implementation of the proposed system {could be}
provided by a chain of superconducting
qubits~\cite{schon,nori,dimitris} in which each qubit operates at
its degeneracy point~\cite{vion}. One of the main reasons for this
choice is related with the remarkable recent advances in
experiments with coupled superconducting
qubits~\cite{yamamoto,plantenberg,ashhab}. The fabrication of
chains of $N\sim50$ Josephson qubits has been achieved in the
laboratory and their coherent operation is a foreseeable
possibility~\cite{experiment}.

The remainder of the {paper} is organized as follows. In
Sec.~\ref{system} we describe the system we use either to obtain a
perfect state transfer or to generate multipartite entanglement of
GHZ form. In Sec.~\ref{protocols} we explain {in details
these two protocols} while in Sec.~\ref{noise} the effects of
static disorder, decay and decoherence on these schemes are
studied. Finally, in Sec.~\ref{remarks} we summarize our results.

\section{The system and the information-flux approach}
\label{system}

The system we analyze is an open spin-chain of $N$ qubits, whose
Hamiltonian reads
\begin{equation}
\label{modelloZZ} \hat{{\cal
H}}=\sum^{N-1}_{i=1}J_{i}\hat{Z}_{i}\hat{Z}_{i+1}+\sum^N_{i=1}B_i\hat{X}_i,
\end{equation}
where $J_i$ is the coupling strength of the pairwise interaction
between qubit $i$ and $i+1$ and $B_{i}$ is a local magnetic field
on qubit $i$. In our notation, $\hat{X},\,\hat{Y}$ and $\hat{Z}$
denote the $x,\,y$ and $z$-Pauli matrix, respectively. The model
in Eq.~(\ref{modelloZZ}) describes, {for instance,} a chain of
interacting superconducting qubits, each at its degeneracy
point~\cite{schon,vion,dimitris}. Here, in order to keep our
discussion on the most general level, free from a specific setup,
we interpret $\hat{\cal H}$ as a generic spin-chain model.

An important remark is due. Frequently one faces the case
of dynamics ruled by spin-preserving Hamiltonian models ({\it
i.e.} Hamiltonians that commute with the total spin operator of
the system). By assuming appropriate boundary conditions, it is
then convenient to diagonalize the coupling {Hamiltonian} by means
of a sequence comprising Wigner-Jordan, Fourier, and Bogoliubov
transformations~\cite{lieb}. In our case, $\hat{\cal H}$
does not preserve the total number of excited spins, so that 
the mutual coupling among subspaces labeled by different numbers of spin-excitation 
has to be considered. Rather than applying techniques for the exact solution of Eq.~(\ref{modelloZZ})~\cite{lieb}, we propose to tackle the evolution of the system by
means of an {\it information-flux approach} (IFA) which is
specifically designed for multi-spin
interactions~\cite{informationflux}. Our method does not rely on
the explicit analysis of the chain's spectrum and enables us to
gather an intuitive picture of the dynamics at hand.

On a formal level, the IFA requires the time-evolved form of
specific operators $\hat{\tilde{O}}_i$ in the Heisenberg picture,
{\it i.e.} $\hat{\tilde{O}}_i(t)=e^{i\hat{\cal
H}t}\hat{O}_ie^{-i\hat{\cal H}t}$ (here $O=X,Y,Z$ and $i=1,..,N$). 
This allows {one to understand the dependence
of} $\hat{\tilde{O}}_i$ on any $\hat{O}_j$ and to design the set
of coupling strengths $\{J_i\}$ in such a way that it becomes possible to
 {\it drive} a
desired evolution by means of engineered quantum
interference~\cite{informationflux}. Our task here is the
transmission of quantum information from the first qubit to the
last one in the chain. This requires the study of $\hat{\tilde{O}}_N(t)$'s, which
can be decomposed into the operator-basis built out of all possible
tensorial products of $\{\hat{X}_i,\hat{Y}_i,\hat{Z}_i\}$.
Therefore, we can write
$\minisand{\Psi_0}{\hat{\tilde{O}}_N(t)}{\Psi_0}=\sum_{O'=X,Y,Z,I}{\cal
I}^{OO'}(t)\minisand{\phi_0}{\hat{O}'_1}{\phi_0}$, where
$\ket{\Psi_0}=\ket{\phi_0}_1\otimes\ket{\psi_0}_{2..N}$ is the
initial state of the whole chain, $\ket{\phi_0}$ is the initial
state of the first qubit only and $\ket{\psi_0}_{2..N}$ is the
state of the rest of the chain. The coefficient ${\cal
I}^{OO'}(t)$ is defined as the information flux at time $t$ from
$\hat{O}'_1$ to $\hat{O}_N$~\cite{informationflux}. It is easy to
see that if our system achieves $|{\cal I}^{OO'}|=1$ when $O=O'$,
we have perfect $1\rightarrow{N}$ state transfer.

To give an immediate picture of IFA, we discuss here a simple but
yet helpful application of the method. We consider a three-qubit
open chain whose Hamiltonian reads $\hat{{\cal
H}}_E=J(\hat{X}_1\hat{X}_2+\hat{Y}_1\hat{Y}_2+\hat{X}_2\hat{X}_3+\hat{Y}_2\hat{Y}_3)$.
By solving the corresponding Schr\"odinger equation, it is well
known that perfect state transfer from the first to the third
qubit is achieved in this case when the initial state of second
and third qubit is $\ket{00}_{23}$ and waiting a time
$t=\pi/(2\sqrt{2} J)$~\cite{cambridge}. By analyzing the evolution
of the operator $\hat{X}_3$, we find
$\hat{\tilde{X}}_3(t)=-\sin^2(\sqrt{2}Jt)\hat{X}_1\hat{Z}_2\hat{Z}_3-(1/\sqrt{2})\sin(2\sqrt{2}Jt)\hat{Y}_2\hat{Z}_3+\cos^2(\sqrt{2}Jt)\hat{X}_3$.
Similar results are obtained by looking at the evolution of
$\hat{Y}_3$ and $\hat{Z}_3$. Therefore, the information flux from
$\hat{X}_{1}$ to $\hat{X}_3$ is ${\cal
I}^{XX}=-\sin^2(\sqrt{2}Jt)_{2,3}\!{\minisand{00}{\hat{Z}_2\hat{Z}_3}{00}}_{2,3}=-\sin^2(\sqrt{2}Jt)$,
whose modulus is $1$ at $t=\pi/(2 \sqrt{2} J)$. We have thus
obtained a unit information flux ({\it i.e.} perfect state
transfer) at the time predicted by the standard approach. Our
simple example shows the basic features of IFA {\it at work}.

\section{The protocols}
\label{protocols}

\subsection{Perfect state transfer}
\label{PST}

\begin{figure}[b]
\centerline{{\bf (a)}}
\vskip0.3cm
\centerline{\psfig{figure=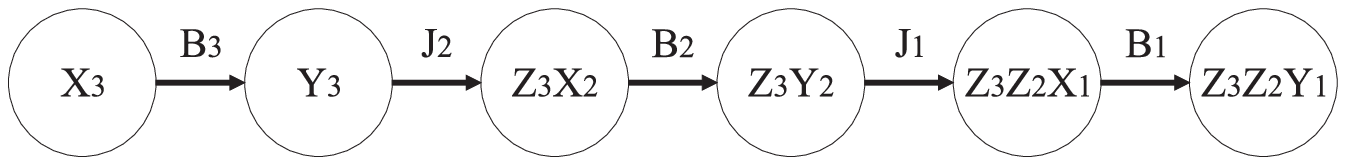,width=8cm}}
\vskip0.3cm
\centerline{{\bf (b)}\hskip3.85cm{\bf (c)}}
\centerline{\psfig{figure=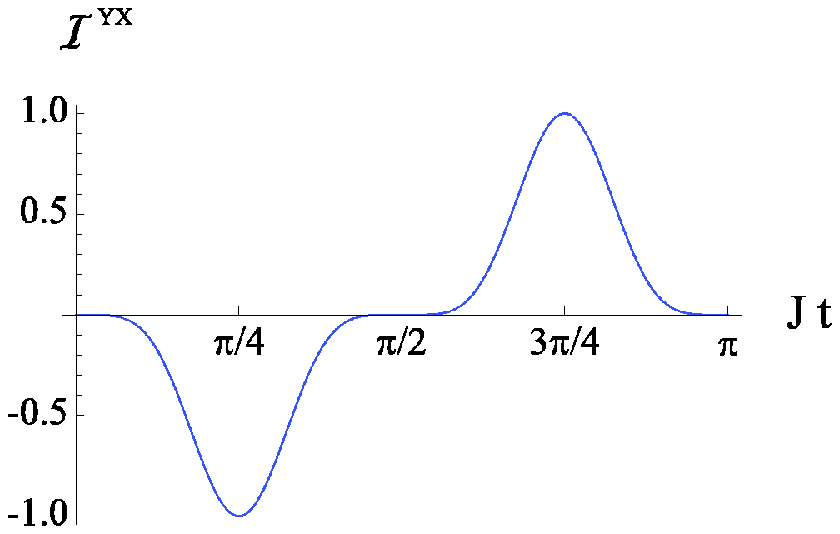,height=2.5cm}
\hskip0.3cm\psfig{figure=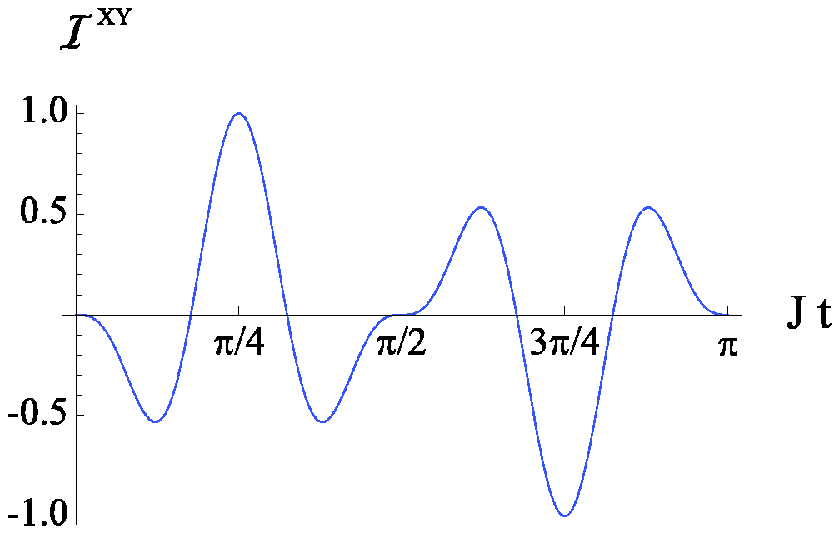,height=2.5cm}}
\caption{{\bf (a)}: Oriented graph describing the way the operators entering  Eqs.~(\ref{evoloper}) are related. The operator in each circle (a node) gives rise to its nearest neighbors under commutation with $\hat{{\cal{H}}}_C$ in Eq.~(\ref{modelloXX}), for $N=3$. The oriented edges connect such nodes. The corresponding coefficients are also shown and an outgoing (ingoing) edge with respect to a node implies a $+$ $(-)$ sign. {\bf (b)}: Information flux from $\hat{Y}_1$ to $\hat{X}_N$ against the dimensionless interaction time $Jt$, for $N=3$ and $J_i=\pm J\sqrt{4i(N-i)}$, $B_i=J\sqrt{(2i-1)(2N-2i+1)}$ in Eq.~(\ref{modelloXX}). Qubits $2$ and $3$ are prepared in $\ket{00}_{23}$. {\bf (c)}: Information flux from $\hat{X}_1$ to $\hat{Y}_N$, for the same conditions as in panel {\bf (b)}.}
\label{flux}
\end{figure}

In order to properly understand our scheme for perfect state
transfer, {it is useful} to consider the Hamiltonian
\begin{equation}
\label{modelloXX} \hat{{\cal
{H}}}_C=\sum^{N-1}_{i=1}J_{i}\hat{X}_{i}\hat{X}_{i+1}+\sum^N_{i=1}B_i\hat{Z}_i.
\end{equation}
By analyzing the evolution of $\hat{X}_N$ and $\hat{Y}_N$ (in the Heisenberg picture) under the action of $ \hat{{\cal{H}}}_C$, we obtain that $\hat{\tilde{X}}_N(t)$ and $\hat{\tilde{Y}}_N(t)$ can be written as
\begin{equation}
\label{evoloper}
\begin{split}
\hat{\tilde{X}}_N(t)&=\mu_1(t)\hat{X}_N+\mu_2(t)\hat{Y}_N+\mu_3(t)\hat{X}_{N-1}\hat{Z}_N+\\
&+\mu_4(t)\hat{Y}_{N-1}\hat{Z}_N+\cdot\cdot+\mu_{2N-1}(t)\hat{X}_1\hat{Z}_2\cdot\cdot\hat{Z}_N+\\
&+\mu_{2N}(t)\hat{Y}_1\hat{Z}_2\cdot\cdot\hat{Z}_N,\\
\hat{\tilde{Y}}_N(t)&=\nu_1(t)\hat{X}_N+\nu_2(t)\hat{Y}_N+\nu_3(t)\hat{X}_{N-1}\hat{Z}_N+\\
&+\nu_4(t)\hat{Y}_{N-1}\hat{Z}_N+\cdot\cdot+\nu_{2N-1}(t)\hat{X}_1\hat{Z}_2\cdot\cdot\hat{Z}_N+\\
&+\nu_{2N}(t)\hat{Y}_1\hat{Z}_2\cdot\cdot\hat{Z}_N.
\end{split}
\end{equation}
The time-dependent coefficients $\mu_i(t)$ and $\nu_i(t)$ are functions of $J_i$ and $B_i$. Their explicit form is too cumbersome to be presented here. However, one can give a graphical picture of the way the operators entering Eqs.~(\ref{evoloper}) are inter-related in terms of oriented graphs, as done in Ref.~\cite{informationfluxijqi}. In particular, it is easy to check that such graphs are linear, in our case. An example, for $N=3$, is given in Fig.~\ref{flux} {\bf (a)}. By explicitly analyzing $\mu_i(t)$ and $\nu_i(t)$, one recognizes that the information flux between $\hat{Y}_1$ ($\hat{X}_1$) and $\hat{X}_N$ ($\hat{Y}_N$) follows the same behavior as the information flux $\hat{X}_1\rightarrow\hat{X}_{2N}$ ($\hat{Y}_1\rightarrow\hat{Y}_{2N}$) in an open $2N$-qubit chain ruled by
\begin{equation}
\label{Hcambridge} 
\hat{\cal H}_{eq}=\sum^{2N-1}_{j=1}J^{eq}_{j}(\hat{X}_j\hat{X}_{j+1}+\hat{Y}_j\hat{Y}_{j+1})
\end{equation}
with $J^{eq}_{j}=B_{\frac{j+1}{2}}$ ($J^{eq}_{j}=J_{j/2}$) for odd
(even) $j$. It is well known that perfect state transfer is
achievable in a spin chain governed by the Hamiltonian in
Eq.~(\ref{Hcambridge}). In particular, if the parameters in
Eq.~(\ref{modelloXX}) follow the pattern $J_i=\pm J\sqrt{4i(N-i)}$ (the choice of the signs needs to be consistent throughout the chain) and
$B_i=J\sqrt{(2i-1)(2N-2i+1)}$~\cite{cambridge} and the initial
state of all the qubits but the first one is $\ket{0}$, a unit (in modulus)
information flux from $\hat{Y}_1$ ($\hat{X}_1$) to $\hat{X}_N$ ($\hat{Y}_N$) is
found at the rescaled dimensionless time $Jt=\pi/4$. The information fluxes $\hat{Y}_1\rightarrow\hat{X}_N$ and $\hat{X}_1\rightarrow\hat{Y}_N$, for $N=3$, are shown in Figs.~\ref{flux} {\bf (b)} and ~\ref{flux} {\bf (c)}, respectively.
This suggests that, by simply applying a single-qubit operator on the first spin
before the evolution under the action of $\hat{{\cal {H}}}_C$ in
Eq.~(\ref{modelloXX}) (the single-qubit operation being required
in order to rotate the operator-basis of the first qubit and
obtain unit flux between homonymous operators), perfect state
transfer can be obtained. Under this point of view, our approach is quite non-conventional. In fact, instead of using a 
formal map between models (\ref{modelloXX}) and (\ref{Hcambridge}), we gather our analytic results and design the optimal protocol 
in virtue of a quantitative analogy between the evolution of specific sets of operators as driven by the two interaction Hamiltonians 
we consider. We remark that this is possible solely because of the power of IFA. Hamiltonian~(\ref{modelloXX}) can be
easily mapped onto the one in Eq.~(\ref{modelloZZ}) and perfect
state transfer thus achieved with a further change of basis for
all the qubits in the chain that transforms
$\ket{0}_i\rightarrow\ket{+}_i$. Let us go into the details of the protocol. We prepare the first
qubit {in} the state
$\ket{\psi}_1=\alpha\ket{0}_1+\beta\ket{1}_1$ and apply the
following recipe for perfect state-transfer:
\begin{itemize}
\item{\underline{First step}}: The operator
$T_1\otimes{H}_1\otimes{T}_1$ is applied to the first qubit of the
chain, where
$T_1=\ket{0}_1\!\bra{0}+e^{i\frac{\pi}{2}}\ket{1}_1\!\bra{1}$ and
$H_1=(\sigma_x+\sigma_z)/\sqrt{2}$ is a Hadamard
gate~\cite{nielsenchuang}. \item{\underline{Second step}}: The
chain evolves under the action of $\hat{\cal H}$ for a time
$Jt=\pi/4$.
\end{itemize}
In this way, the state of the last qubit will be $\ket{\psi}_N$,
while the rest of the chain in the tensorial product state
$\ket{++..+}_{1..N-1}$, thus achieving perfect state transfer.

The first step is required to cope with both the necessary basis
changes (the {first} one to map the Hamiltonian $\hat{\cal
H}$ {onto} $\hat{{\cal {H}}}_C$ and the {second} one
to obtain unit fluxes between homonymous operators) and can be
performed off-line. This corresponds to carrying out only the second
step but using the initial state
$\miniket{\tilde{\psi}}_1=[({\alpha+i\beta})\ket{0}_1+({\beta+i\alpha})\ket{1}_1]/{\sqrt{2}}$.
Alternatively, the change of basis can also be performed off-line
at the end of the evolution driven by $\hat{\cal H}$. The initialization of the register in the tensorial 
product of $\ket{+}_i$ states can be obtained by cooling a thermal state of the qubit system to their own $\ket{0}_i$ state (thus achieving $\ket{00\cdot\cdot0}_{1..N}$, which is the standard initial state in quantum state-transfer protocols~\cite{cambridge,unmodulated}) and then applying single-qubit Hadamard gates. The same task can also be achieved by using a collective external potential resonant with the transition between the single-qubit levels, waiting enough time for the system to relax to the ground state and then applying the Hadamard gates.

\subsection{Generation of multipartite entanglement}

\label{GHZ}

\begin{figure}
\psfig{figure=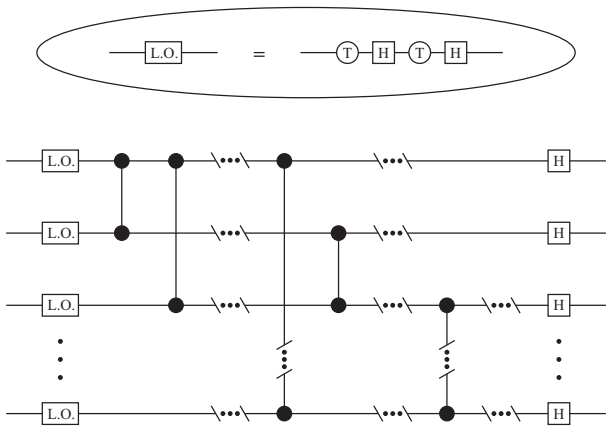,width=8cm}
\caption{Equivalent quantum circuit for a spin-chain that evolves
according to $e^{-i\hat{\cal H}\frac{\pi}{4J}}$. The inset shows
the explicit composition of the local operators before the
inter-qubit interaction structure. We show the symbols for T, H
and controlled-phase operations. For clarity of presentation, we have omitted to show a set of swap gates (performing mirror-inversion) at the end of the circuit.} \label{equiv}
\end{figure}

Interesting features arise from a deeper analysis of the model
encompassed by $\hat{\cal H}$. Indeed, the same parameter pattern
addressed up until now can be used so as to generate an N-qubit
GHZ state
$\ket{GHZ}_{12..N}=(\bigotimes^N_{i=1}\ket{0}_i-i\bigotimes^N_{i=1}\ket{1}_i)/\sqrt{2}$.
If the initial state of every qubit is set to be $\ket{0}$, after
the action of $\hat{\cal H}$ for a time $Jt=\pi/4$, the system
will be in state $\ket{GHZ}_{12..N}$. This is understood by
considering the fact that the evolution
$e^{-i\frac{\pi}{4J}\hat{\cal H}}$ can be decomposed into the
equivalent quantum circuit shown in Fig.~\ref{equiv}~\cite{analogouscircuit}. It is then
straightforward to check that a GHZ state of the above mentioned
form is the result of the computation described there.

It is very important to stress that our scheme to generate
entanglement does not require any pre-built entanglement resource.
Multipartite entanglement is generated in the chain only as a
result of spin-non-preserving dynamics. An analogous situation has
recently been addressed in~\cite{matryoshka}.

\section{Effects of static disorder, decay and decoherence}
\label{noise}

We address the effects of static disorder in the pattern of
coupling strengths which is required for the optimal performance
of the protocols. This {is useful in a general situation,
and} {particularly necessary in the case of superconducting
qubits.} Although control over lithographic techniques being used
in the fabrication of superconducting qubits {is constantly
being} improved, it is reasonable to expect that the elements of a
long chain of effective spins would not be identical and the
inter-qubit couplings will correspondingly be affected. We also
consider the {influences of} Markovian dissipation and
phase-decoherence on the state transfer and the creation of a
multipartite GHZ state. In order to account for the usual finite
spatial environmental correlation length that frequently affects
solid-state systems, we address both the case of uncorrelated
individual baths being coupled to the spins in a chain as well as
the case of the same bosonic environment affecting multiple spins
simultaneously. Our study reveals not only a considerable
resilience of the protocols at hand but also points out some
interesting features that highlight the counter-intuitive nature
of noise effects in purely quantum dynamics. It is worth stressing that, 
as we study a coupling model with a novel pattern of interaction strengths, which is 
different from the one given in~\cite{cambridge}, an investigation about noise effects 
is quite significant and represents a due step in the full characterization of the 
scheme we study. Such a program is in line with previous studies on fluctuation and noise in state transfer 
protocols, which can be found in Refs.~\cite{dech,referee}.

\subsection{Perfect state transfer}
\label{PSTnoise}

We start by analyzing the effects of static disorder in the
coupling-strength pattern for the state transfer protocol. Our
model for such imperfections is that of a distribution of coupling
strengths along the chain according to~\cite{dech}
\begin{equation}
\label{modeldisorder}
\begin{aligned}
J_i\rightarrow{\cal J}_i&=J_i[1+\delta(1-2{r}_i)],\\
B_i\rightarrow{\cal B}_i&={B}_i[1+\delta(1-2{s}_i)]
\end{aligned}
\end{equation}
with $\delta$ the `strength' of the disorder (typically a few
percent of $J$) and $r_{i},s_{i}\in[0,1]$ that decide whether the imperfect
parameter is larger or smaller than the ideal value. In our
study, we take $r_{i}$'s and $s_{i}$'s as random numbers following
a given probability distribution. Clearly, the choice of such a
distribution should be made consistent with a given experimental
situation. Here, in order to gather a simple idea of what the
effect of static disorder is, a uniform distribution is chosen.
Intuitively, this case should depict a sort of `worst case'
scenario for the influences of disorder. We show that, for values
of $\delta$ which are within the current accuracy of lithographic
fabrication of superconducting qubits, this class of imperfections
does not significantly affect the performance of quantum state
transfer in our chain.

The figure of merit that we choose in order to test the
performance of the protocol affected by disorder is the state
fidelity between the logical input qubit and the state of the
$N^{th}$ physical qubit of the chain at time $t=\pi/4J$.
Indeed, while it is reasonable to expect that disorder would
influence the state of the rest of the chain, leaving it in  a
state different from the expected
$\bigotimes^{N-1}_{i=1}\ket{+}_i$, this is irrelevant in a
state-transfer problem. Clearly, in a real physical situation, we
have no available information on the exact value of the disordered
coupling strengths' pattern. This implies that a faithful measure
of the quality of our protocol would be rather given by the
average of state fidelity over a large sample of disorder
configurations (from now on called {\it runs}). That is, if we
indicate as ${\cal F}_{1N}(\delta,\{r_i,s_i\}_k)=\langle{ist}|{\varrho^{\delta,\{r_i,s_i\}_k}_N}|ist\rangle$ the state fidelity corresponding to a given run labelled by the integer
$k=1,2,..,M$ (with $\ket{ist}$ is the ideal state to transfer and $\varrho^{\delta,\{r_i,s_i\}_k}_{N}$ the reduced density matrix of the last qubit of the chain), we will consider the average value
\begin{equation}
{\cal F}^{av}_{1N}=\frac{1}{M}\sum^M_{k=1}{\cal
F}_{1N}(\delta,\{r_{i},s_{i}\}_k).
\end{equation}
The cut-off $M$ in the estimation of the average is taken large
enough to guarantee that no appreciable differences in ${\cal
F}^{av}_{1N}$ are observed if a larger $M$ is taken.

\begin{figure}[b]
\centerline{{\bf (a)}\hskip3.85cm{\bf (b)}}
\centerline{\psfig{figure=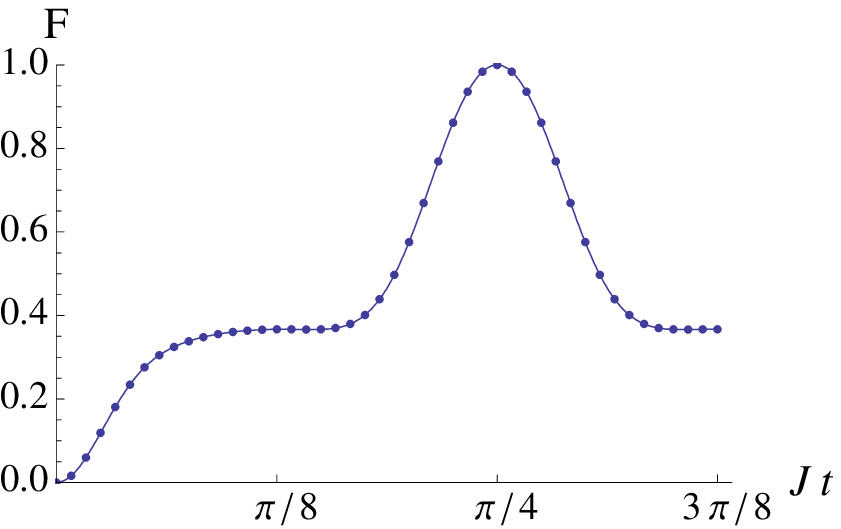,height=2.5cm}
\hskip0.3cm\psfig{figure=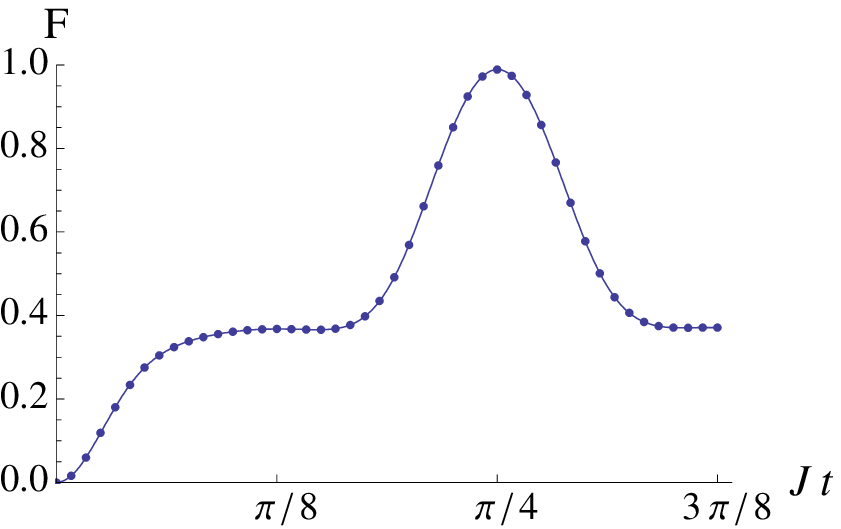,height=2.5cm}}
\caption{{\bf (a)}: State fidelity against dimensionless time $Jt$
with no disorder and $N=7$; {\bf (b)}: Same with $\delta=0.05$,
corresponding to a $5\%$ maximum deviation of the disordered
parameters from the ideal values.} \label{disordine}
\end{figure}

We first consider a state-dependent situation based on an analogy
with existing studies on quantum state transfer. We will shortly
demonstrate that the state fidelity of the process is very weakly
dependent on the input state being considered. However, in order
to fix the ideas, we consider the arbitrarily chosen input state
$\ket{-}_{1}=(\ket{0}_1-\ket{1}_1)/\sqrt{2}$. As a trade off between
required computational power and length of the chain, we have
examined the case of $N=7$ qubits throughout the paper (where not
stated differently). Although the results presented in our study depend on the length of the chain system 
being studied, we have checked that such a dependence is very weak and our findings can 
be taken as faithful indications of the behavior of even longer chains. As the maximum of fidelity is expected for
$Jt=\pi/4$, we restrict our observation to a temporal window
centered in this value. In Fig.~\ref{disordine}, we compare the
ideal and the disordered situation for a strength $\delta/J=5\%$,
which is a typical  in state of the art value~\cite{pc}.

Evidently, the influence of disorder is negligible, at least for
the operating conditions considered above: not only is the height
of the fidelity peak preserved almost perfectly, but also such
peak occurs precisely at the re-scaled time that is expected under
ideal performances. This is important, experimentally, as it
proves that our state of ignorance over the exact value of the
disordered coupling strengths does not require the fine adjustment
of the time at which the logical qubit to be transferred should be
collected.

\begin{figure}[t]
\centerline{{\bf (a)}\hskip4.2cm{\bf (b)}}
\psfig{figure=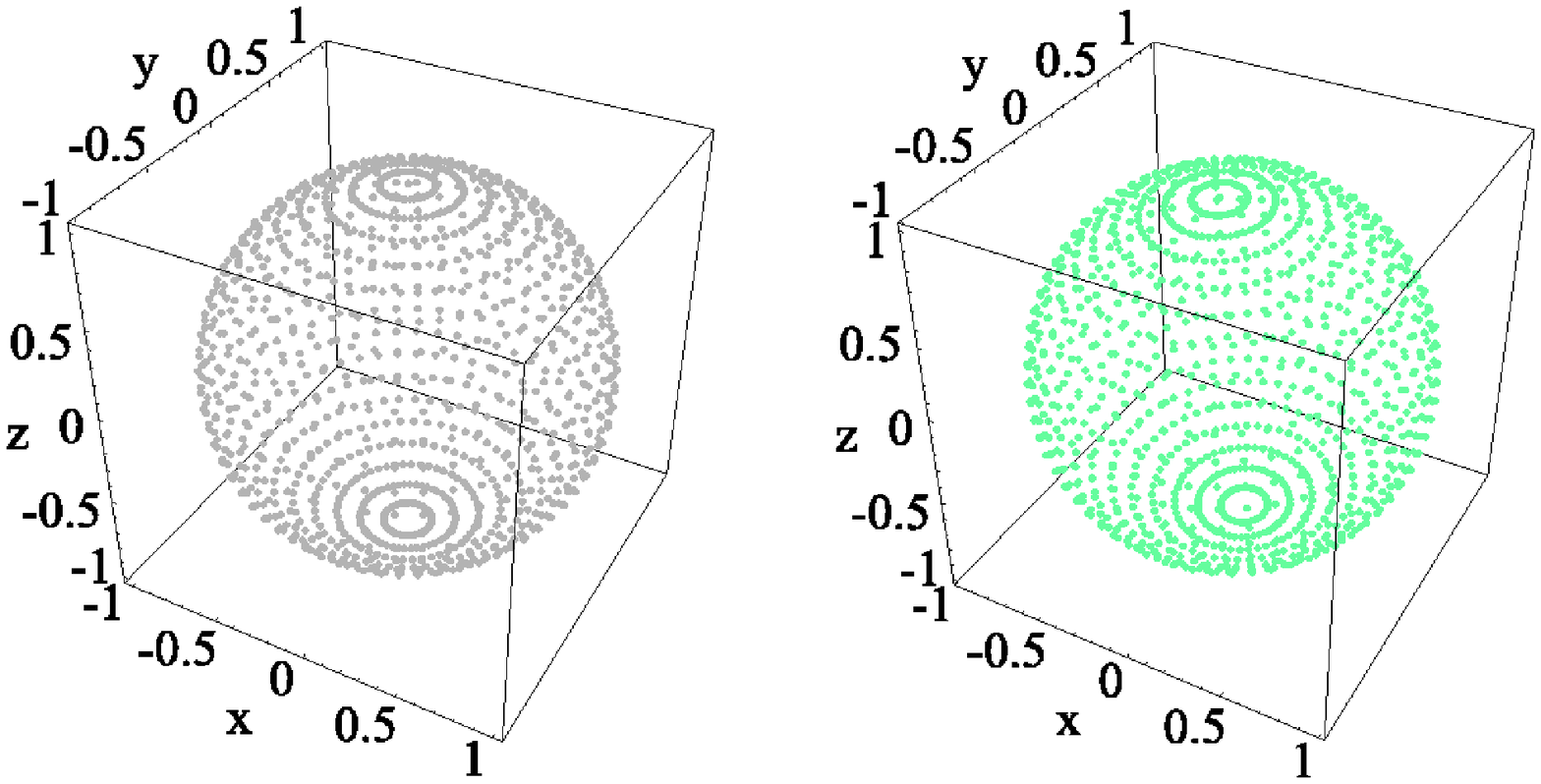,height=4.5cm}
\caption{{\bf (a)}: Bloch sphere of the logical input qubit for
various choices of $\alpha$. {\bf (b)}: Reconstructed Bloch sphere
showing the state of the $N^{th}$ qubit for a disordered chain of
$N=7$ and $\delta/J=5\%$ at $Jt=\pi/4$.}. \label{QPT0QPT5}
\end{figure}

As a step forward in the analysis of the protocol, we now remove
any dependence on the logical state to be transferred. We take a
`black box' perspective under which the problem at hand is not
different from that of characterizing an unknown operation (the
transfer channel with disorder) with respect to an ideal one which
would map the logical input state onto the same state at the
$N^{th}$ physical qubit, leaving the logical input state
unchanged. Such a characterization is efficiently performed by
using quantum process tomography (QPT) techniques~\cite{qpt,mark},
which has proven to be invaluably useful in many experimental
contexts~\cite{experQPT}. QPT is a useful tools for the evaluation
of the global closeness between the ideal and disordered transfer
channel. By studying the way the black box affects the transfer of
the four {\it probe states}
$\{\ket{0},\ket{1},\ket{+},\ket{+_y}\}$, we can reconstruct the
so-called {\it process matrix} $\chi_{dis}$ of the disordered
process. This contains all the relevant information regarding the
process that a logical qubit undergoes in being transferred from
the first to the last physical qubit. We refer to existing
literature~\cite{qpt,mark} for further mathematical details and
just mention that the QPT approach can also be computationally
beneficial as the averaged state fidelity can be calculated out of
the process fidelity
\begin{equation}
F_p={\rm Tr}(\chi_{id}\chi_{dis})
\end{equation}
with $\chi_{id}$ the process matrix of the logical identity
operation that is applied to the logical qubit to transfer in the
perfect case. In the operator basis
$\{\hat{\openone},\hat{X},-i\hat{Y},\hat{Z}\}$, this reads
\begin{equation}
\chi_{id}=
\begin{pmatrix}
1&0&0&0\\
0&0&0&0\\
0&0&0&0\\
0&0&0&0
\end{pmatrix}.
\end{equation}
QPT allows the reconstruction of the effective Kraus operators of
the disordered transfer channel and, from these, one can easily
get the output density matrix for any logical input state. In
turn, by using the correspondence between density matrix and Bloch
vector, we can visualize the effect of the disorder directly on
the Bloch sphere of the transferred state, thus getting a general
overview of the behavior of the protocol. Given the results in
Fig.~\ref{disordine}, the output Bloch sphere with disorder
strength $\delta/J=5\%$ at $Jt=\pi/4$ should be very close to the
Bloch sphere of a pure qubit state. This is indeed the case, as
shown in Fig.~\ref{QPT0QPT5}, where panel {\bf (a)} shows the
Bloch sphere for the ideal protocol, while panel {\bf (b)}
addresses the case of the disordered pattern corresponding to the
worst fidelity among $200$ runs. The two processes are evidently
quite close to each other, the most evident difference being a
slight unwanted rotation of the sphere's poles around the $x$
axis. We have checked that the rotation angle is an increasing
function of the disorder's strength. No shrinking of the sphere
for the disordered case is evident, showing that coherence and
amplitude of the logical qubit are almost perfectly preserved. The
matrix of the disordered process is given by
\begin{equation}
\label{diprocmat} \chi_{dis}=
\begin{pmatrix}
0.987&0.003+0.022i&0&0\\
0.003-0.022i&0.001&0&0\\
0&0&0.008&0.003\\
0&0&0.003&0.005
\end{pmatrix},
\end{equation}
which corresponds to a state fidelity of $0.991$. The state
fidelity averaged over the whole set of runs is as large as
$0.995$. The closeness of this number to the worst-case fidelity
shows that the disordered state transfer process depends only very
weakly on the pattern of disorder. On the other hand, the evident 
isotropy of the reconstructed Bloch sphere justifies an arbitrary
choice of the input state for our explicit analysis. We can
therefore confidently affirm that, within the range of chain's length 
we have studied, static disorder does not spoil
the process we have designed for perfect state transfer across a
chain. The average fidelity of the process stays largely above $2/3$, the highest value for a classical transmission of a
state~\cite{horo}, up until values of $\delta/J\sim30\%$.

We now consider the effects of dissipation and decoherence
on the protocol. We assume conditions of weak coupling
of the spins of the chain with the environment, so that a spin-boson
approach can be retained and the consequent non-unitary dynamics
can be studied within the context of a Lindbladian approach or,
equivalently, an operator-sum representation of the
evolution~\cite{Preskill}. By choosing this second perspective, as
it will clearly appear in what follows, we model our study along
the lines of a Monte Carlo simulation. The environmental action
over the chain is identified in the effects of both amplitude and
phase damping channels. First, we address the case of $N$
individual baths, one for each spin of the chain. This assumption
is physically motivated by considering that each Josephson charge
qubit has its own voltage gate and can thus suffer {\it
individual} electromagnetic fluctuations. Then in order to account
for the finite spatial correlation length of a Markovian
environment, which may occur in solid-state systems, we consider a
bosonic environment affecting more than a single spin
simultaneously (as can be the case for a bath of background
phonons arising from the common substrate onto which the charge
qubits are placed). In the following, we show a considerable robustness
of the state transfer protocol against both these sources of
imperfection and the appearance of some counter-intuitive features
that can be related to the phenomenon of stochastic
resonances~\cite{susanamartin}.

In an operator-sum representation, the evolution of an initial
state $\varrho_c$ of the whole chain under the effect of an
environment is given by
\begin{equation}
\varrho_c(t)=\sum_{\mu}\hat{K}^{\mu}(t)\varrho_c\hat{K}^{\mu\dag}(t)
\end{equation}
with $\{\hat{K}^\mu(t)\}$ a set of time-dependent Kraus operators,
having the structure of a tensorial product of single-spin
operators and such that
$\sum_{\mu}\hat{K}^{\mu\dag}(t)\hat{K}^\mu(t)=\openone$~\cite{nielsenchuang,Preskill}.
The formal description of a single-spin amplitude damping process
in a bath at finite temperature is described by the
trace-preserving completely positive map given by the following
set of Kraus operators~\cite{nielsenchuang}:
\begin{equation}
\label{genampldamp}
\begin{aligned}
\hat{A}^0_{i}&=\sqrt{p}
\begin{pmatrix}
1&0\\
0&e^{-\frac{\Gamma{t}}{2}}
\end{pmatrix},\,
\hat{A}^1_{i}=\sqrt{p}
\begin{pmatrix}
0&\sqrt{1-e^{-\Gamma{t}}}\\
0&0
\end{pmatrix}\\
\hat{A}^2_{i}&=\sqrt{1-p}
\begin{pmatrix}
e^{-\frac{\Gamma{t}}{2}}&0\\
0&1
\end{pmatrix},\,
\hat{A}^3_{i}=\sqrt{1-p}
\begin{pmatrix}
0&0\\
\sqrt{1-e^{-\Gamma{t}}}&0
\end{pmatrix}
\end{aligned}
\end{equation}
with $p=(\nbar+1)/(2\nbar+1)$ and $\nbar$ the average phonon
number of each bath. For a phase damping channel, on the other
hand, we have the following set~\cite{nielsenchuang}
\begin{equation}
\label{phasedamp} \hat{D}^0_i=\sqrt{\frac{1+e^{-\gamma{t}}}
{2}}\hat{\openone}_i,\,\hat{D}^1_i=\sqrt{\frac{1-e^{-\gamma{t}}}{2}}\hat{Z}_i.
\end{equation}
\begin{figure}[b]
\centerline{\hskip1cm{\bf (a)}\hskip3.85cm{\bf (b)}}
\centerline{\psfig{figure=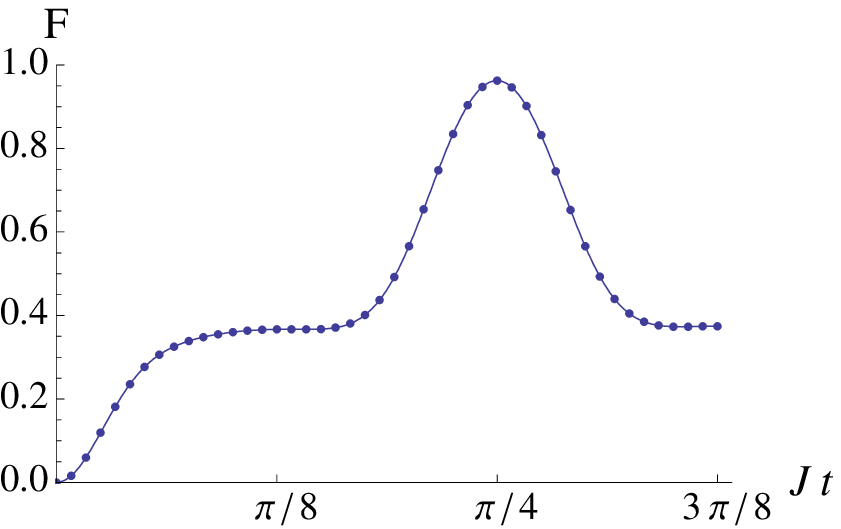,height=3.25cm}
\hskip0.cm\psfig{figure=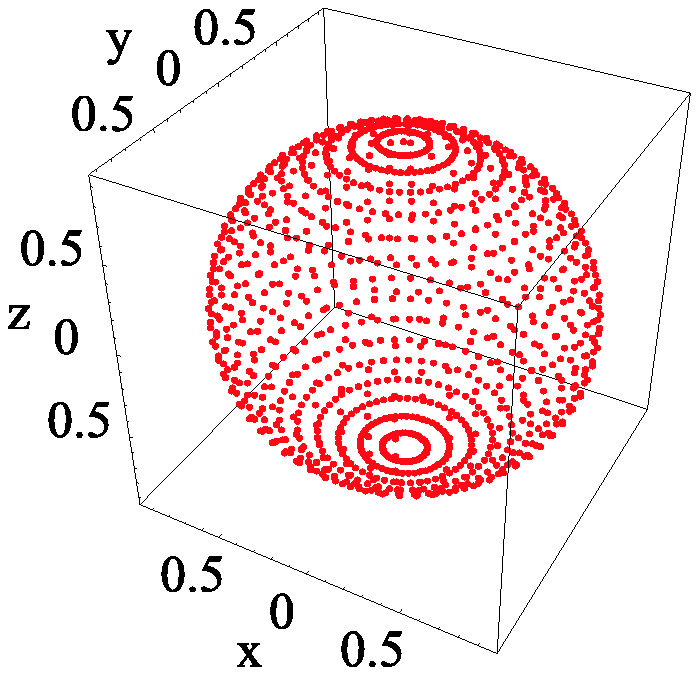,height=3.25cm}}
\caption{{\bf (a)}: Fidelity ${\cal F}^{av}_{1N}$ for an input
state $\ket{-}_1$ against $Jt$ with $\delta/J=5\%$,
$\Gamma/J=0.5$, $\gamma/J=0.2$, $\nbar=0.01$ and $N=7$. The
average fidelity is calculated over a set of $200$ disordered
patterns. {\bf (b)}: Bloch sphere of the output qubit evaluated
with QPT for the same parameters assumed in panel {\bf (a)}.}
\label{deco}
\end{figure}
In the above equations, $\Gamma$ and $\gamma$ are the rates of
amplitude and phase damping, respectively. On the other hand,
clearly, the unitary dynamics determined by the coupling
Hamiltonian in Eq.~(\ref{modelloXX}) leads to
$\varrho_c(t)=e^{-i\hat{\cal H}t}\varrho_c{e}^{i\hat{\cal H}t}$.
Our approach is to intersperse a unitary evolution lasting for a
time interval $\Delta{t}$ with non-unitary dynamics. $\Delta{t}$ is taken randomly according to the general recipe for 
Quantum Monte Carlo simulations (see Ref.~\cite{QMC}, for instance). Moreover, we 
randomly determine the set of Kraus operator to
apply and the specific spin being affected. The
disorder's pattern is kept as fixed for the duration of one of
such simulated dynamics. In virtue of the weak dependence of state
fidelity on $\delta$, a random choice of the disorder
configuration will be enough. The state fidelity resulting from
the simulated dynamics has then to be averaged over a collection
of noise-occurrence patterns (runs), which guarantees the faithful unraveling of the open quantum dynamics~\cite{QMC}. 
For our numerical study, whose results are presented in  Fig.~\ref{deco} {\bf (a)}, we have taken
the input state $\ket{-}_1$ with $\delta/J=5\%$, $\gamma/J=0.5$,
$\Gamma/J=0.2$ and $\nbar=0.01$ in a chain of $7$ qubits and a
sample of $200$ runs~\cite{commento}. Evidently, the state
fidelity is close to $0.95$, which is an excellent result. Again,
the peak of fidelity is obtained at the expected value $Jt=\pi/4$,
although the effects of noise and decoherence, altering the
pattern of quantum interference that is at the basis of a state
transfer process~\cite{unmodulated,espanso}, could also have
resulted in a shift of the peak. A more complete picture comes
from the analysis by means of QTP. As before, we have chosen a
random configuration of disordered coupling strengths and have
then evaluated the effects of decoherence and dissipation on a
generic input state, obtaining the output Bloch sphere in
Fig.~\ref{deco} {\bf (b)}. The effects of noise seem to result in
just a small uniform shrinking of the Bloch sphere. The average fidelity of the process is $0.959$.

\begin{figure}
\psfig{figure=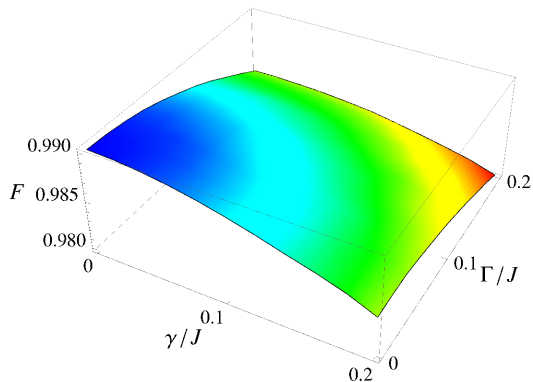,height=5cm} 
\caption{Fidelity ${\cal F}^{av}_{1N}$ versus $\gamma/J$ and $\Gamma/J$ for $Jt=\pi/4$,
$\delta/J=5\%$, $\nbar=0.01$ and $N=6$.}
\label{3D}
\end{figure}

\begin{figure}[b]
\psfig{figure=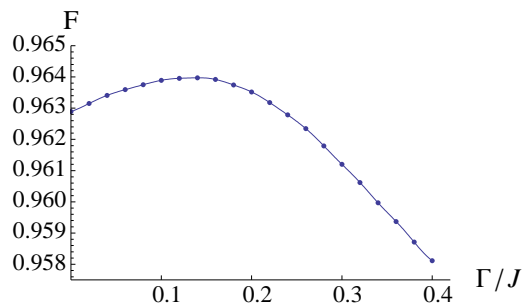,height=4cm} \caption{Fidelity ${\cal
F}^{av}_{1N}$ with $N=6$ at $Jt=\pi/4$, $\delta/J=5\%$,
$\gamma/J=0.2$, $\nbar=0.01$ against $\Gamma/J$. The transmission
fidelity is maximal for an optimal value of the environmental
noise and is the result of the average over many noisy and disordered runs. 
The fact that the efficiency of the protocol depends
non-monotonically on the noise strength can be viewed as a form of
stochastic resonance.}
\label{susana}
\end{figure}

In order to evaluate the relative {\it weight} of the two noisy
channels over the state fidelity, we have plotted its maximum
against $\gamma$ and $\Gamma$ and for one randomly generated
disorder pattern (Fig.~\ref{3D}). In this case, we needed to
reduce the number of qubits in the chain to $N=6$ and increase the
number of runs to obtain a better plot. Moreover, in order to 
exclude any dependence on the input state, we have averaged the state 
fidelity over a large sample of random pure input states. In virtue of the 
almost perfect insensitivity of the protocol to the specific instance of 
input state, this step turned out to be rather conservative.
The most relevant feature appearing from this study is
that, for a fixed value of $\gamma$, the state fidelity has a
(weak) non-monotonic convex behavior against $\Gamma$. As this
feature is not well visible from Fig.~\ref{3D}, we have considered
a bidimensional projection of this plot along the cut
$\gamma=0.2J$ in Fig.~\ref{susana}, where the interaction time is
fixed at $Jt=\pi/4$. The existence of a small peak of fidelity for
increasing value of the dissipation rate can be interpreted following
the lines of Ref.~\cite{susanamartin}, where the phenomenon of
{\it quantum stochastic resonances} has been extended to a 
multiparticle setting and invoked as the mechanism behind
counter-intuitive features of benchmarks such as entanglement
against the influences of noise~\cite{susanamartin,briegel}.
We observe that a {\it not too weak} dissipative mechanism alters the
occurrence of quantum interference so as to improve the
performances of the state transfer protocol. This effect obviously
disappears as soon as the dissipation is so strong for the system
to cope with it in a coherent way. The non-monotonic transmission
fidelity provides us with a new way to characterize the response
of a system which differs from the figures of merit used
in~\cite{susanamartin}, where dynamical properties (such as the 
chain's magnetization) or information theoretic ones (such as the 
mutual information) have been used as a measure of global correlations.
Our study goes along the lines of previous results by Mancini and
Bowen~\cite{mancinibowen}, where the rate of transmission of
information through a class of quantum channels is shown to
increase with the external noise. The results of our investigation
motivate further studies of stochastic resonance phenomena in the
context of information transmission through quantum channels.

\begin{figure}
\psfig{figure=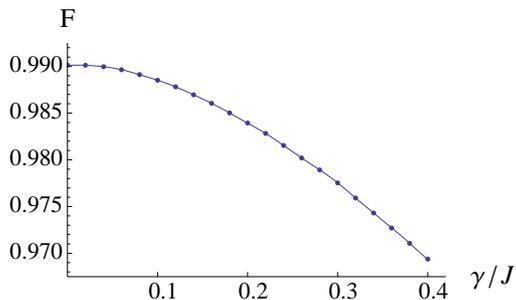,height=4cm} 
\caption{Fidelity ${\cal F}^{av}_{1N}$ for a chain interacting with collective baths
addressing two qubits simultaneously with $Jt=\pi/4$,
$\delta/J=5\%$ and $N=6$ against $\gamma/J$ and in absence of dissipation.}
\label{sharedplot}
\end{figure}

We conclude our analysis of the effects of realistic imperfections
on the proposed protocol for quantum state transfer by addressing
the case of baths having a longer correlation length, so as to
give rise to non-localized environmental effects. The analysis so
far has been done by considering every qubit interacting with its
own independent bath. As explained before, the substrate where our
hypothetical superconducting chain is fabricated could provide a
mechanism for correlated noise at longer haul. In this case the
qubits can be considered, in first instance, to interact with
shared baths. Here, in order to provide an intuitive idea of the
general behavior of the protocol with respect to this situation,
we consider the simplest case in which we have a collection of
independent baths, each affecting a pair of spins at the same
time. In order to model the evolution of the system in this
scenario we use the operator-sum representation corresponding to
the set of Kraus operators for two-qubit collective phase damping
noise (see Ref.~\cite{shared}, for instance)
\begin{equation}
\begin{aligned}
\hat{D}^c_0&=
\begin{pmatrix}
e^{-\gamma{t}/2}&0&0&0\\
0&1&0&0\\
0&0&1&0\\
0&0&0&e^{-\gamma{t}/2}
\end{pmatrix},\\
\hat{D}^c_1&=\sqrt{1-e^{-\gamma{t}}}
\begin{pmatrix}
1&0&0&0\\
0&0&0&0\\
0&0&0&0\\
0&0&0&-e^{-\gamma{t}}
\end{pmatrix},\\
\hat{D}^c_2&=({1-e^{-\gamma{t}}})
\begin{pmatrix}
0&0&0&0\\
0&0&0&0\\
0&0&0&0\\
0&0&0&\sqrt{1+e^{-\gamma{t}}}
\end{pmatrix}.
\end{aligned}
\end{equation}
Due to our choice of two qubit-addressing collective bath, here we
consider a chain of six qubits. Clearly, the analysis can be
extended to any number of qubits being simultaneously addressed.
We have obtained the plot in
Fig.~\ref{sharedplot}. 
In the presence of phase decoherence alone, no stochastic resonance 
effects are observable. The fidelity decreases monotonically as a function
of $\gamma/J$, which agrees with all previous literature~\cite{sr}, 
where the emergence of stochastic resonance phenomena was shown to require
the presence of noise forms that couple transversely to the
localized basis states.

Although the Hamiltonian model we have considered does not enjoy
particularly symmetries with respect to the coupling with the
collective baths, so that no genuine phase-damping decoherence
free subspace can be singled out, the resilience of the protocol
is evident. The results are comparable to the case of individual
environment attached to each spin and we have checked that this
feature does not depend on the number of qubits being affected at
the same time by a given bath~\cite{remark}.

\subsection{Generation of multipartite entanglement}
\label{GHZnoise}

We now pass to a brief analysis of the proposed protocol for GHZ-state
generation along the lines described in the previous part of our
study. However, here there is an important technical difference
that has to be stressed: differently from what has been done
above, here we need to consider the global fidelity of chain with
a GHZ state of the form
$\ket{GHZ}_{12..N}=(\bigotimes^N_{i=1}\ket{0}_i-i\bigotimes^N_{i=1}\ket{1}_i)/\sqrt{2}$.
Indeed, we are now interested in the generation of a multipartite
entangled state where every qubit will be involved. We have
evaluated the evolution of the system in the noisy scenario. We
take advantage of the analysis performed in the previous Section
and in order to avoid unnecessary redundancy, we present here 
the results obtained including static disorder, dissipation and
phase decoherence, without considering them individually.

\begin{figure}[b]
\centerline{{\bf (a)}\hskip3.75cm{\bf (b)}}
\centerline{\psfig{figure=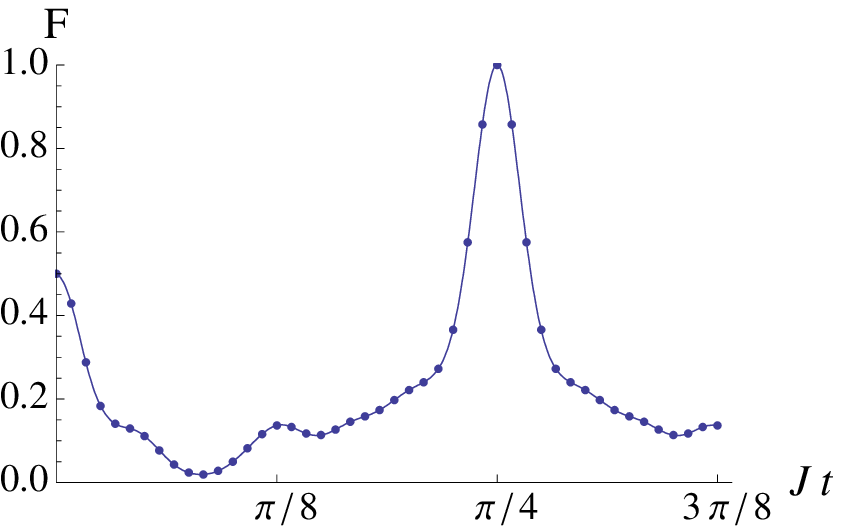,height=2.5cm}\hskip0.3cm\psfig{figure=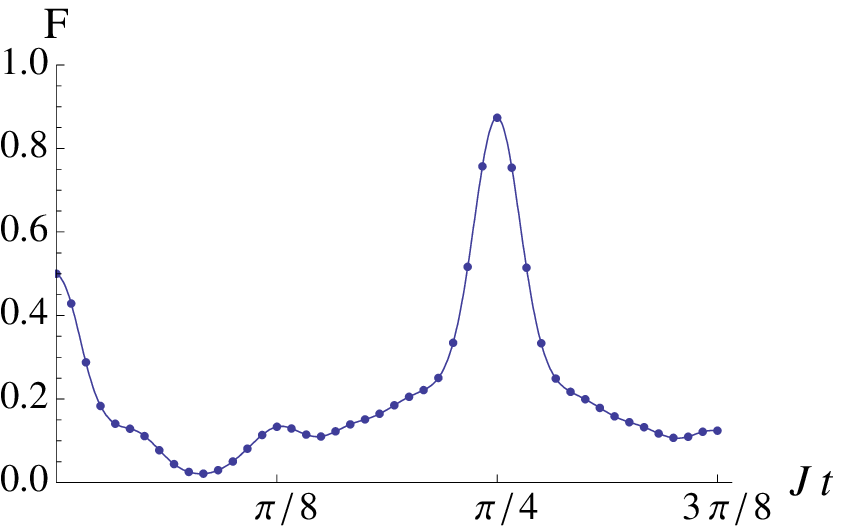,height=2.5cm}} \caption{{\bf
(a)}: State fidelity for GHZ-state generation against
dimensionless time $Jt$ in the ideal case and $N=7$; {\bf (b)}:
Same with disorder $\delta/J=5\%$, $\gamma/J=0.5$, $\Gamma/J=0.2$
and $\nbar=0.01$.}
\label{ghz}
\end{figure}

Fig.~\ref{ghz} shows the fidelity against dimensionless time $Jt$
in the ideal case [panel {\bf (a)}] and for the same values of
disorder, decoherence and dissipation rates and temperature used
previously [panel {\bf (b)}]. The chain we have considered is
seven-qubit long and the state fidelity has been averaged over 200
different runs. Although the performance of the protocol seems to
be inferior to the one corresponding to quantum state transfer, we
should keep in mind that here we are using a global figure of
merit. Moreover, despite the fragility of GHZ-like entanglement to
various forms of noisy channels~\cite{noiseGHZ}, the resilience of the imperfect
multipartite-entanglement generation protocol is still quite
satisfactory (maximum fidelity larger than
$0.88$).

\section{Concluding Remarks}
\label{remarks}

We have used information flux methods in order to design a 
protocol for quantum state transfer in a finite, open chain of
qubits. The protocol is based on a nontrivial map between the
algebra of angular momenta an a control-limited Ising-like
Hamiltonian with additional local magnetic fields. The dynamics
described by such a coupling is rich enough to provide an
effective way of performing a multipartite entangling gate among
the elements of the register, which produces GHZ-like
entanglement. Both these protocols have been {analyzed} in
terms of their resilience {against certain realistic
sources} of disorder and decoherence {and were found} to be
rather robust. {Moreover,} our study has {uncovered}
an unexpected relation between state-transfer fidelity and quantum
stochastic resonances, thus {giving} our
pragmatically-oriented work a more fundamental character. Our
investigation contributes to the affirmation of information
flux-based methods in the design of multipartite systems in
control-limited scenarios and to the ubiquitous nature of
stochastic resonance mechanisms, two points which certainly
deserve further studies.

\acknowledgments 

We thank F.~Ciccarello, M.~S.~Kim, M.~B.~Plenio, and M.~S.~Tame for
fruitful discussions. We acknowledge support from the EPSRC, the
QIPIRC, and the FP6-IP QAP (``Qubit Applications''). M.P. was
supported by The Leverhulme Trust (Grant No. ECF/40157) and the Bridging Fund from Queen's University Belfast. D.I.T. was supported by EPSRC (EP/D065305/1).


\begin{thebibliography}{99}
\bibitem{commento} For weakly coupled spins $J\sim{1}$GHz is experimentally reasonable in charge qubits (the coupling between two qubits in Ref.~\cite{yamamoto} was around $10$GHz). This implies the rather pessimistic values of $\Gamma^{-1}\sim{0.2}$ns and $\gamma^{-1}\sim{0.5}$ns, which are one order of magnitude smaller than the decoherence time estimated experimentally in~\cite{yamamoto} for two coupled charge qubits.
\bibitem{remark} {As an additional remark, we briefly mention about the possibility of an error in the timing of the process. The protocol requires the extraction of the logical state after the precise time $t=\pi/(4J)$, which corresponds to the maximum of the state fidelity. This is obviously equivalent to switching off the mutual inter-spin interaction at time $\pi/(4J)$ and extracting the state at the $N^{th}$ spin in the chain. As we have already said, a realistic choice for $J$ is in the range of $1$GHz (it can be larger) while the temporal window within which, typically, the state fidelity is larger than $0.9$ (an arbitrary and rather conservative choice) is as large as $J\Delta{t}\simeq{0.2}$. This leads to $\Delta{t}$ being typically in the range of $0.1$ns, which can be perfectly resolved experimentally~\cite{yamamoto}.}
\end{thebibliography}
\end{document}